# Kinematics of the HH 111 jet from STIS spectra [1]

A. C. Raga[1], A. Noriega-Crespo[2], B. Reipurth[3], P. M. Garnavich[4], S. Heathcote[5], K. H. Böhm[6], S. Curiel[7]

## ABSTRACT

We present a long-slit spectrum of the Herbig-Haro object HH 111, obtained with STIS on the Hubble Space Telescope. This spectrum has a spectral resolution of $\approx 50$ km s$^{-1}$, so that it gives a good picture of the kinematical properties of the observed object at very high angular resolution. We find that some of the knots along the jet are associated with sudden drops in the radial velocity (modulus), confirming that the emission from the knots is formed in shocks. We interpret the observed position-velocity diagrams in terms of a model of a jet from a variable source, and we attempt to carry out a reconstruction of the ejection velocity variability necessary for reproducing the observed kinematical structure.

*Subject headings:* ISM: Herbig-Haro objects — ISM: jets and outflows — ISM: kinematics and dynamics — ISM: individual (HH 111) — shock waves

[1]Instituto de Ciencias Nucleares, UNAM, Apdo. Postal 70-543, 04510 D. F., México, email: raga@astroscu.unam.mx

[2]SIRTF Science Center, California Institute of Technology, IPAC 100-22, Pasadena, CA 91125, email: alberto@ipac.caltech.edu

[3]Institute for Astronomy, University of Hawaii, 2680 Woodlawn Drive, Honolulu, HI 96822

[4]Center for Astrophysics, Department of Physics, 225 Nieuwland Science Hall, University of Notre Dame, Notre Dame, IN 46556-5670

[5]Cerro Tololo Inter-American Observatory, NOAO, Casilla 603, La Serena, Chile

[6]Department of Astronomy, FM-20, University of Washington, Seattle, WA 98195

[7]Instituto de Astronomía, UNAM, Ap. 70-264, 04510 D. F., México

## 1.  Introduction

Following the discovery paper of Reipurth (1989, which included images and long slit spectroscopy), the spectacular HH 111 system has received a substantial amount of attention. Spectroscopic and Fabry-Perot studies describing its excitation and the kinematics (Noriega-Crespo et al. 1993; Morse et al. 1993a, b; Reipurth et al. 1997a; Rosado et al. 1999), imaging (Reipurth et al. 1997a, b, the former paper describing HST images) and proper motion measurements (Reipurth et al. 1992; Hartigan et al. 2001 using HST images) have been carried out at optical wavelengths. At infrared wavelengths, narrow-band imaging (Gredel & Reipurth 1993; Reipurth et al. 1999, using HST), spectroscopy (Davis et al. 2001) and proper motion studies (Coppin et al. 1998) have been made. Finally, radio continuum (Reipurth et al. 1999) and CO observations (Cernicharo & Reipurth 1996; Nagar et al. 1997; Hatchell et al. 1999) of this system have also been performed. Also, a number of theoretical studies with a specific application to the HH 111 system have been carried out (e. g., Raga & Noriega-Crespo 1993; Bacciotti et al. 1995; Völker et al. 1999; Lee et al. 2000), and several of the observational papers listed above also contain models of this object.

In the present paper, we discuss a STIS (HST) spectrum of the HH 111 jet. The only other STIS observations of an HH jet which have been carried out so far are the ones of the DG Tauri "small-scale jet" described by Bacciotti et al. (2000), so that our observations provide an unique, high spatial resolution view of the kinematics of a "classical" HH jet.

## 2.  The observations

We have obtained spectroscopic observations of the HH 111 jet with STIS covering the [6295-6867] Å wavelength range. A $0\overset{''}{.}1$ slit width was used, giving a spectral resolution of $\approx 50$ km s$^{-1}$ FWHM, and a dispersion of $0\overset{''}{.}554$ Å pix$^{-1}$. The spatial sampling was of $0\overset{''}{.}05$ pix$^{-1}$.

The observations were obtained on December 9th and 16th, 2000 and consisted of seven exposures over two visits, with an average exposure time of 2400 seconds each. Each exposure was split into three to remove cosmic rays. The exposures were dithered along the slit to help to remove hot pixels.

The spectra were reduced using standard HST calibration steps and then shifted to create a single image with an equivalent exposure of 4.7 hours. On the second visit, only one guide star could be fine-locked, causing a small drift of $\approx 0\overset{''}{.}1$ (mostly along the slit direction) during the exposures. As each exposure was split into 3 parts, the shift within each of these parts was only of a fraction of a pixel, and most of the effect of the shift was then corrected by making small offsets of the resulting spectra along the direction of the slit.

The slit was oriented at PA=-83°.94 (coinciding with the jet axis), and covered the region along the HH 111 jet from a distance of $12''$ to $62''$ from the outflow source. Actually, only the bright region from knots E to L was detected in the [O I] 6300/6363, Hα and [S II] 6717/6731 lines. This is consistent with previously obtained spectra of this region of HH 111 (see, e. g., Morse et al. 1993).

## 3. Position-velocity diagrams and line ratios

In Figure 1, we show a red [S II] archival HST image of HH 111 (rotated so that the outflow axis is approximately parallel to the ordinate) on which the position of the spectrograph slit of our STIS observations is superposed. The position-velocity diagrams obtained for the [O I] 6300, Hα, and the shifted and co-added red [S II] lines are shown in the three other frames of Figure 1. The radial velocities are given with respect to the systemic velocity (for which we have chosen an LSR velocity of $+23$ km s$^{-1}$, following Reipurth 1989).

Knots F, H and J are clearly seen in the three position-velocity (PV) diagrams. Knot L is also seen, but only in Hα. The region between knots F and H (which includes knot $G_1$) is detected in [S II] and [O I], but is barely seen in the Hα PV diagram.

The three PV diagrams appear to share the same kinematical characteristics (see figure 1). As a function of increasing distance from the source, we first see a region with a relatively constant radial velocity $v_r \sim -60$ km s$^{-1}$, ending in a sharp velocity drop at the position of knot F (at $\approx 28''$ from the source). A faint ramp of increasing (i. e., more negative) velocity vs. position is then seen, ending in a second sharp drop at the position of knot H (at $\approx 33''$ from the source). At larger distances from the source, only the emission of knots J (with a $v_r \sim -50$ km s$^{-1}$ radial velocity) and L (with $v_r \sim -100$ km s$^{-1}$, detected only in Hα) is seen.

We have also attempted to analyze the line ratios (obtained from the line intensities integrated across the line profiles) as a function of position along the slit. We find that the signal-to-noise ratio of the line fluxes is mostly too low to obtain significant line ratios. The best results are obtained for the $r_{S\,II}$ =[S II] 6716/6731 line ratio in the regions around knots F and H, which we show in Figure 2. In this Figure, we see that in knot F we have apparently real fluctuations in the $r_{S\,II} = 0.4 \rightarrow 1$ range, which correspond to electron densities in the $n_e = 500 \to 10^4$ cm$^{-3}$ range. In knot H, the red [S II] line ratio is more constant as a function of distance from the source, and remains mostly within the $r_{\mathrm{S\,II}} = 0.5 \to 0.7$ range, corresponding to a $n_e = 1500 \to 4000$ cm$^{-3}$ electron density range. Notably, knots F and H both show a quite dramatic increase in $r_{\mathrm{S\,II}}$ (corresponding to a sharp drop in electron density) towards the leading edges of the knots. These sharp drops in $n_e$ might be consistent with a shock travelling into a lower density, upstream medium.

As the [S II] PV diagram has the best overall signal-to-noise ratio, we discuss it in more detail. A contour plot of the region containing knots F, G and H is shown in Figure 3. We find that the FWHM of the line has a basically position-independent value of $\approx 65$ km s$^{-1}$, except for knot F, which shows a larger width of $\approx 76$ km s$^{-1}$. This lack of variability of the FWHM as a function of position is a direct result of the limited spectral resolution of our observations (see §2).

More interesting are the results obtained by fitting a quadratic curve to the line peaks to determine the central radial velocity of the line profiles as a function of position. The results of this fitting procedure are also given in Figure 3, and show the general kinematical characteristics described above.

The more remarkable feature is the drop from $v_r = -62$ to $v_r = -23$ km s$^{-1}$ which occurs at the position of knot F. The spatial scale of this drop is unresolved in our observations, in which it occurs over a distance along the slit of 1 pixel (0″.05), corresponding to a spatial scale of $\approx 4 \times 10^{14}$ cm (assuming a distance of 460 pc to HH 111). This very sharp velocity drop is quite convincing evidence for the presence of a shock in knot F.

A less abrupt radial velocity drop is seen at the leading edge of knot H. At a distance $x = 28''$ from the source, the radial velocity (modulus) drops from $v_r = -58$ to $v_r = -36$ km s$^{-1}$ over 0″.2 ($\approx 3 \times 10^{15}$ cm).

The regions in between the knots have a more or less continuous variation of radial velocity vs. position. An interpretation of these variations in terms of a model of a jet from a time-dependent source is discussed in the following section.

## 4. The position-dependent jet velocity and a time-dependent ejection

Considering a $\phi = 11°$ orientation angle of the outflow axis with respect to the plane of the sky (see Reipurth et al. 1992; Hartigan et al. 2001), we can deproject the distance and the radial velocity, in order to obtain the true velocity of the jet as a function of distance from the source. The obtained results are given in Figure 4, in which we show the deprojected radial velocity vs. position obtained from the [S II] PV diagram (also see figure 3).

For $x < 1.9 \times 10^{17}$cm, we see a ramp of decreasing jet velocity $v_j$ vs. position. Then, we have the sharp velocity drop occurring at knot F. For $1.9 \times 10^{17}$cm$< x < 2.15 \times 10^{17}$cm we have a ramp of increasing $v_j$. For $x > 2.15 \times 10^{17}$ cm, we again have a ramp of decreasing $v_j$ vs. $x$, which ends in a second sharp drop in $v_j$ at the position of knot H. At larger distances from the source, the [S II] emission is barely detected in our observations (see figures 1 and 3).

Using the $v_j$ vs. $x$ determined from our [S II] PV diagram, we can determine the dynamical time $t = -x/v_j$ at which the fluid parcels would have been ejected from the source ($t = 0$ corresponding to the time at which the observation was made) if they travelled ballistically along the jet beam. The results of this calculation are shown in the bottom frame of Figure 4.

If one had a straight, ballistic jet, one would expect the modulus of the dynamical time $t$ to increase monotonically as a function of distance $x$ from the source (as required by the "no trajectory crossing" condition satisfied by a fluid). In the case of a jet with internal working surfaces produced by an ejection velocity variability, one would see monotonic ramps of increasing $t$ vs. $x$, with discontinuous increases in $t$ at the positions of the working surfaces.

Such a structure is present for $x < 1.9 \times 10^{17}$cm, where we see an increasing $t$ vs. $x$ ramp which ends at a sharp edge in which the modulus of $t$ increases from a value of $\approx 200$ yr to $|t| \approx 500$ yr. A similar structure is seen in the $2.15 \times 10^{17}$ cm$< x < 2.33 \times 10^{17}$ cm region, in which we again have a ramp of increasing $|t|$ vs. $x$ ending in a sharp rise in $|t|$.

The $1.9 \times 10^{17}$ cm$< x < 2.15 \times 10^{17}$ cm region is more problematic. In this region, we see a noisy but generally decreasing $|t|$ vs. $x$ dependence. As we have stated above, such a dependence does not occur in a straight, ballistic jet. Therefore, we conclude that the emission observed in this region cannot be coming from ballistic material ejected from the outflow source.

Let us now consider the $x < 1.9 \times 10^{17}$cm and the $2.15 \times 10^{17}$ cm$< x < 2.33 \times 10^{17}$ cm regions, which in principle can be interpreted in terms of a ballistic flow. We have fitted the $v_j$ vs. $x$ dependence of these two regions with the almost straight, monotonic curves (labeled A and B) in Figure 4, in order not to be distracted by the substantial apparent noise in the determined velocities. From these two curves, we can calculate the (smooth) dynamical time $t$ as a function of $x$, and then plot $v_j$ as a function of $t$.

The result of this exercise is shown in Figure 5, which can be interpreted as giving the history of the time-dependent ejection velocity from the source as a function of time $t$. In this plot, the $t < -195$ yr segment of the curve corresponds to region B of the jet beam (with $x > 2.15 \times 10^{17}$ cm), and the $t > -190$ yr segment corresponds to region A (with $x < 1.9 \times 10^{17}$ cm, see figures 4 and 5). In a really surprising way, we find that these two physically disconnected regions of the jet beam map into two contiguous segments in the $v_j$ vs. $t$ plot, which are separated by a very sharp drop of $\approx 40$ km s$^{-1}$ in $v_j$. This sharp drop occurs over a period smaller than $\sim 5$ yr, though it is not possible to determine its precise duration from our rather noisy data.

If we take the deduced $v_j$ vs. $t$ dependence seriously, we would have the following situation. Approximately 210 yr ago (at $t \approx -195$ yr, see figure 5), the source was ejecting the outflow with a velocity $v_j \approx 360$ km s$^{-1}$, and there was then a very sudden drop of $\approx 40$ km s$^{-1}$ to a $v_j \approx 320$ km s$^{-1}$ ejection velocity. This sudden drop in velocity (corresponding to a $\Delta M \approx 4$ drop in Mach number, for an assumed jet temperature of $\sim 10^4$ K) will start to produce a gap in the jet beam, which opens up as the previously ejected, higher velocity material "runs away" from the later, slower velocity outflow. This gap would correspond to the $1.9 \times 10^{17}$ cm$< x < 2.15 \times 10^{17}$ cm region along the presently observed beam of the HH 111 jet (see figures 3 and 4).

One can only speculate from where the emission observed in this gap could come from. Obvious candidates are the wings of the bow shocks associated with previous knots (for example, extended bow shock wings from knot H). These bow shock wings would correspond to an "entrainment region" around the beam of HH 111, which was indentified kinematically by Reipurth et al. (1997a).

We should point out that a major source of uncertainty in the interpretation of the PV diagrams is that the emission along the HH 111 jet shows excursions of $\sim 0\rlap{.}''3$ to both sides of the outflow axis (see figure 3 and Reipurth et al. 1997a), so that different regions of the cross section of the outflow fall within our $0\rlap{.}''1$ wide slit. This could be a major effect in the production of the low velocity region observed at distances $1.9 \times 10^{17}$ cm$< x < 2.15 \times 10^{17}$ cm from the source (see figure 4).

## 5. Conclusions

Using the HST, we have obtained a STIS spectrum of the HH 111 jet. From this spectrum, we have obtained PV diagrams of the [O I] 6300, H$\alpha$ and the (shifted and co-added) red [S II] lines. This last PV diagram has the best signal-to-noise ratio, so we have employed it to obtain a description of the kinematics of the jet as a function of position along the slit.

We find that at the positions of knots F and H, sharp drops in the radial velocity are present. These sharp drops in velocity at the positions of the knots are in qualitative agreement with what one would expect from an interpretation of the knots as "internal working surfaces" resulting from an ejection velocity time-variability (see, e. g., Raga et al. 1990; de Gouveia dal Pino & Benz 1994).

We find that the regions in between the knots have somewhat more confusing kinematic characteristics. Part of the inter-knot region kinematics can indeed be interpreted in terms of a ballistic jet from a variable source, but other parts clearly do not correspond to material following a ballistic motion.

Using the regions that could be the result of ballistic motions, we have reconstructed the ejection velocity $v_j$ vs. time $t$ that would have produced the observed PV diagrams. We find that the reconstructed time variability has two ramps of increasing $v_j$ vs. $t$, separated by a basically discontinuous drop in the ejection velocity.

This kind of velocity variability is quite different from the one assumed for modelling the knots along HH jets and would result in a jet that develops "gaps" in the beam in part of the inter-knot regions. We then speculate that the wakes filling in these gaps could correspond to the non-ballistic regions observed along the HH 111 jet.

The results obtained from our STIS observations are therefore in partial agreement with the idea that the knots in HH 111 correspond to internal working surfaces along the jet beam. However, the detailed kinematic characteristics of the PV diagram appear to imply that the source time-variability necessary for producing these working surfaces is substantially different from the variabilities that have been explored in the previously studied theoretical models. Therefore, further studies will be necessary to clarify whether or not the scenario that we are proposing for explaining our high angular resolution PV diagrams of HH 111 is actually correct.

Another issue is whether or not the deduced ejection velocity variability is consistent with time-dependent models of jet collimation (Ouyed & Pudritz 1997; Goodson et al. 1999), or with the possibility that the outflow variability could be due to perturbations of an accretion disk due to companions of a binary or multiple outflow source (Reipurth 2000). This point also requires a detailed future study.

Support for this publication was provided by NASA through proposal number GO-08277 from the STScI. The work of A. Raga was supported by the CONACyT 32753-E and 34566-E grants. A. Raga acknowledges support from a Fellowship of the John Simon Guggenheim Foundation. We acknowledge Tom Ray (the referee) for his helpful comments.

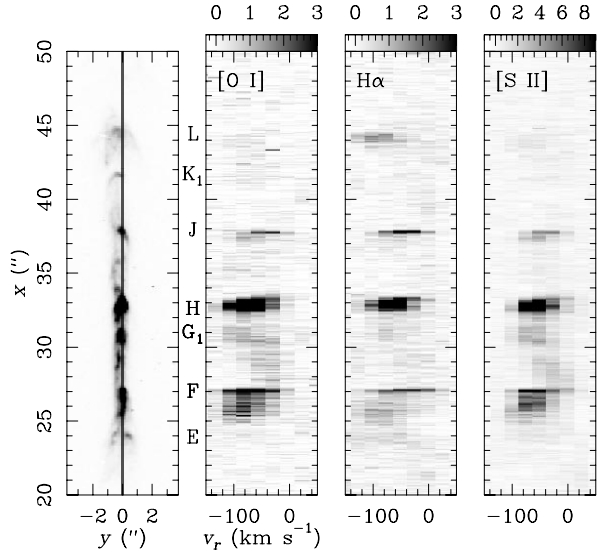

Fig. 1.— From left to right : [S II] HST image, and [O I] 6300, Hα, and combined [S II] (6716+31) PV diagrams obtained from our STIS observations. The [S II] image is shown with a linear greyscale, and the position of the slit is shown on the image (with the thick vertical line). The PV diagrams are also shown with linear greyscales, with the corresponging fluxes given (in $10^{-15}$erg cm$^{-1}$s$^{-1}$ Å$^{-1}$ ($''$)$^{-2}$) by the bars above each plot. The distances are measured along the outflow axis from the approximate position of the source ($x$) and perpendicular to the outflow axis from the center of the spectrograph slit ($y$). The radial velocities are given with respect to the systemic velocity (see the text and Reipurth 1989). Several of the knots along the HH 111 jet are identified.

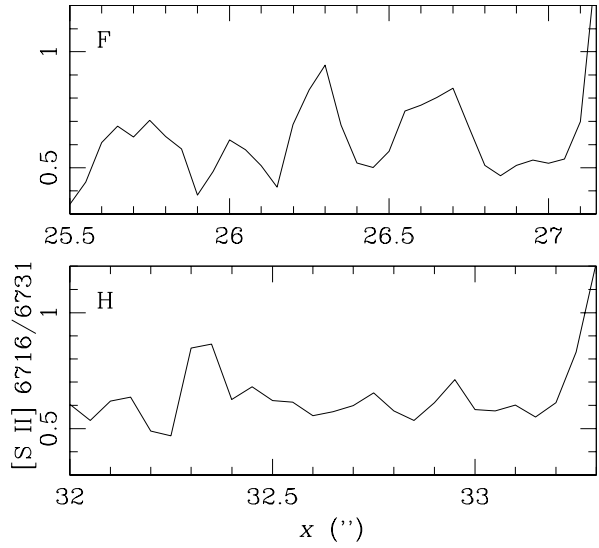

Fig. 2.— [S II] 6716/[S II] 6731 line ratio (obtained from the line intensities integrated over the line profiles) as a function of distance $x$ from the source for the two brightest regions, around knots F (above) and H (below).

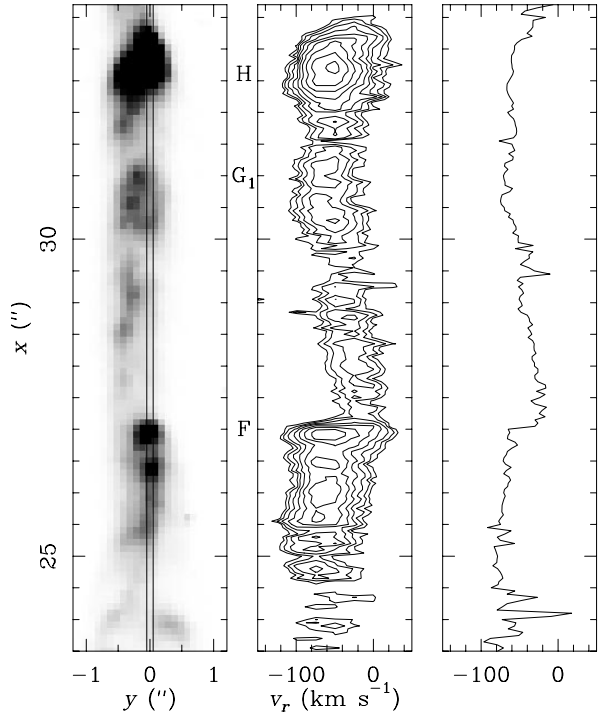

Fig. 3.— [S II] HST image (left), [S II] PV diagram (center) and radial velocity vs. distance (right, determined from fits to the peak of the [S II] emission) of HH 111. The position of the spectrograph slit (two vertical lines) is shown on the [S II] image. The PV diagram is depicted with logarithmic, $\sqrt{2}$ contours.

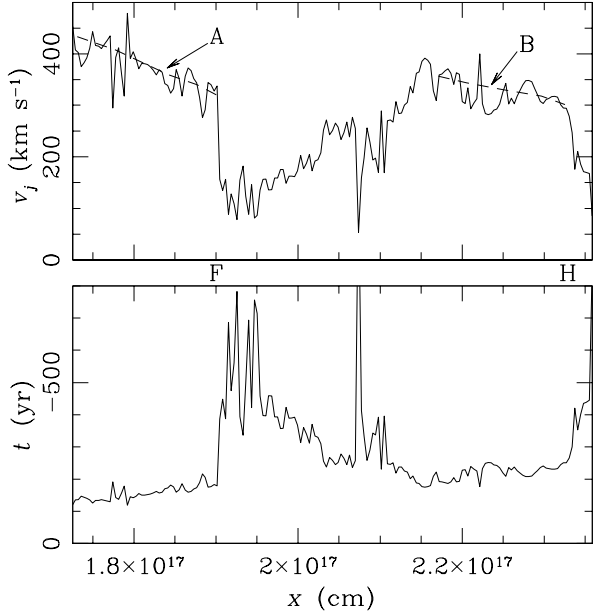

Fig. 4.— Top : Deprojected jet velocity $v_j$ (obtained from the [S II] PV diagram) vs. distance $x$ from the source (calculated assuming a distance of 460 pc to HH 111). Bottom : dynamical time $t = -x/v_j$ computed from the observed $v_j$ vs. $x$ dependence. Two regions of the velocity vs. position dependence of the HH 111 jet can be approximate with the smooth curves (indicated with the dashed lines) labeled A and B. The approximate positions of knots F and H are also indicated.

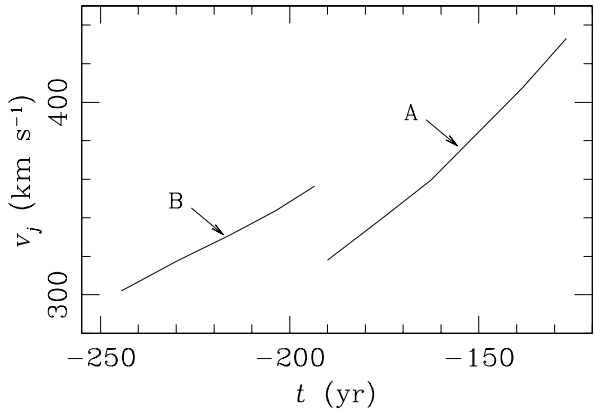

Fig. 5.— Ejection velocity vs. time dependence reconstructed from the two segments of the HH 111 jet shown with dashed lines in Figure 4 (also see the text). The two jet beam segments map onto two increasing $v_j$ vs. $t$ ramps, which are separated by a sharp velocity drop (at $t \approx -195$ yr). These two ramps are labeled with the same letters as the corresponding segments in the $v_j$ vs. $x$ diagram (see figure 4). $t = 0$ corresponds to the time at which the observation was made.